\documentclass[twocolumn]{aastex631}

\usepackage[english]{babel}
\usepackage{longtable}
\usepackage{bbm}
\usepackage{enumitem}
\usepackage{graphicx}
\usepackage{xcolor}
\usepackage{wrapfig}
\usepackage{soul}
\usepackage{physics}
\usepackage{amssymb}

\newcommand\E[1]{{\left\langle #1 \right\rangle}}

\newcommand\correction[1]{{#1}}
\definecolor{KB}{RGB}{0,128,128}

\usepackage[normalem]{ulem}

\newcommand*{\pprob}{\mathrm{P}}
\newcommand*{\gwL}{\mathcal{L}}

\newcommand*{\BinaryParameters}{\vec{\lambda}}
\newcommand*{\AllBinaryParameters}{\{\vec{\lambda}\}}
\newcommand*{\FormationParameters}{\Lambda}
\newcommand*{\gwdetj}{d_j}
\newcommand*{\gwdets}{\{\gwdetj\}}

\newcommand*{\metallicity}{\mathcal{Z}}
\newcommand*{\lumdist}{\mathrm{\ell}}
\newcommand*{\tobs}{T_{\mathrm{obs}}}

\newcommand*\mc{\mathcal{M}_c}
\newcommand*{\fgw}{f_{\mathrm{gw}}}
\newcommand*{\alphaCE}{\alpha_{\mathrm{CE}}}
\newcommand*{\msun}{M_{\odot}}
\newcommand*{\zsun}{\metallicity_{\odot}}

\newcommand*{\citepLVK}{\citep{GW150914-detection}}
\newcommand*{\citepLVKRP}{\citep{LIGO-O3-O3a-RP, LIGO-O3-O3b-RP, LIGO-O3-O3b-NSBH}}
\newcommand*{\citepLVKIsolated}{\citep{Hurley2002,Eldridge2017,Giacobbo2017,Breivik2020,Belczynski2020-EvolutionaryRoads, broekgaarden2021formation,broekgaarden2021impact, posydon, Stevenson2022, st_inference_interp}}
\newcommand*{\citepLVKDynamic}{\citep{RomeroShawGW190521,Fragione_2020,Gayathri2022,Gamba2023Nature}}

\newcommand*{\citepLISAMission}{\citep{LISA-white}}
\newcommand*{\citepLISAGalacticBinaries}{\citep{Nelemans2001,Ruiter2010,Toonen2012TypeIa,Korol2017,Korol2018,Lamberts2019,Breivik2020,Korol2022, Hirai2022, Biscoveanu2023Outspiral, Tang2024}}

\newcommand*{\citepOSG}{\citep{osg07,osg09}}

\newcommand{\AffiliationCCRG}{
  Center for Computational Relativity and Gravitation, 
  Rochester Institute of Technology, 
  Rochester, New York 14623, USA 
}
\newcommand{\AffiliationGSFC}{
  Gravitational Astrophysics Laboratory, 
  NASA Goddard Space Flight Center, 
  Greenbelt, MD 20771, USA 
}
\newcommand{\AffiliationCMU}{
  McWilliams Center for Cosmology and Astrophysics, 
  Physics Department, 
  Carnegie Mellon University, 
  Pittsburgh, PA 15213, USA
}
\newcommand{\AffiliationPrinceton}{
  Department of Astrophysical Sciences, 
  Princeton University, 
  Princeton, NJ 08544, USA 
}

\begin{document}
\title{
       Recovering Injected Astrophysics from the LISA Double White Dwarf Binaries
}
\author[0000-0001-7099-765X]{V. Delfavero}
\affiliation{\AffiliationGSFC}

\author[0000-0001-5228-6598]{Katelyn Breivik}
\affiliation{\AffiliationCMU}

\author[0000-0001-7442-6926]{Sarah Thiele}
\affiliation{\AffiliationPrinceton}

\author{R. O'Shaughnessy}
\affiliation{\AffiliationCCRG}

\author{J. G. Baker}
\affiliation{\AffiliationGSFC}

\date{\today}

\clearpage{}\begin{abstract}
We present the successful recovery of common envelope ejection efficiency
    assumed in a simulated population of double white dwarf binaries
    like those which may be observed by the future LISA mission.
We simulate the formation of double white dwarf binaries by using 
    the COSMIC population synthesis code
    to sample binary formation conditions 
    such as initial mass function, metallicity of star formation,
    initial orbital period, and initial eccentricity.
These binaries are placed in the \textbf{m12i} synthetic Milky-Way-like galaxy,
    and their signal-to-noise ratio for the LISA instrument
    is estimated, considering a Galactic gravitational wave foreground informed by
    the population.
Through the use of Fisher estimates,
    we construct a likelihood function for the measurement error of
    the LISA-bright DWD binaries 
    ($\ge 20$ SNR, $f_{\mathrm{GW}} \ge 5 \mathrm{mHz}$),
    in their gravitational wave frequency ($f_{\mathrm{GW}}$)
    and chirp mass.
By repeating this process for different assumptions of the common envelope ejection efficiency,
    we apply Bayesian hierarchical inference to find the best match
    to an injected astrophysical assumption
    for a fiducial population model.
We conclude that the impact of common envelope ejection efficiency
    on the mass transfer processes involved in double white dwarf formation
    may be statistically relevant in the future observed LISA population,
    and that constraints on binary formation 
    may be found by comparing simulated populations to a future observed population.

\end{abstract}
\clearpage{}

\section{Introduction}
\label{sec:intro}
In the past decade, the LIGO/Virgo/KAGRA collaboration has published
    the first observations of gravitational waves from
    compact binary mergers in the kilo-Hertz frequency band
    \citepLVK{}.
These measurements probe populations of stellar mass
    binary black holes and neutron stars 
    which electromagnetic observations cannot probe
    except by targeting X-ray emission from accreting binaries
    \citep{Fabbiano2006,Remillard_2006,miller-jones_2014, Inoue2022, Fortin2023}.
In the 2030s, the Laser Interferometer Space Antenna (LISA) will
    be the first space-based gravitational wave interferometer
    \citepLISAMission{}.
LISA will be most sensitive in a lower gravitational wave frequency than
    its extant ground-based counterparts
    and will observe detached compact binaries,
    such as the Double White Dwarf (DWD)
    population in The Milky Way
    \citepLISAGalacticBinaries{} .

The LISA Galactic DWD binaries will probe isolated binary evolution
    in the lower end of the stellar mass spectrum where most stars reside.
Several million DWDs in The Milky Way and local group are expected to 
    radiate gravitational waves in the LISA band, with the majority having overlapping 
    frequencies that can be observed as a Galactic gravitational-wave 
    foreground \citep[e.g.][]{Benacquista2006, Breivik2020b, Staelens2024,Hofman2024}.
In addition to the Galactic foreground, is expected that LISA will discover tens-of-thousands of individually 
    resolved DWDs in The Milky Way \citep[e.g.][]{Nelemans2001,Nissanke2012,Tang2024} 
    and the local group \citep[e.g.][]{Korol2018b, Roebber2020, Korol2020, Keim2023}.
In addition to isolated binaries,
    LISA will discover DWDs that have been formed either dynamically
    in environments like globular clusters 
    \citep[][]{Willems2007,Kremer2018},
    or higher-multiplicity systems in the field of The Milky Way
    \citep[e.g.][]{2023ASPC..534..275O}.
However, the Type Ia supernova rate from merging DWDs in globular clusters
    may be as low as $10^{-7} yr^{-1}$
    in an environment like The Milky Way
    \citep{Kremer2020,Kremer2021},
    compared to a $10^{-3}yr^{-1}$ estimate for the Type Ia rates
    of Milky-Way-like galaxies \citep[][]{Frohmaier2018}.
\correction{
Furthermore, the fraction of WD progenitor stars 
    formed in stellar triples is measured to make up 
}
    $20\%$ or less of the stellar population \citep[e.g.][]{2023ASPC..534..275O}.
Thus, the majority of resolved DWDs discovered by LISA are 
    expected to have been born in isolated binary systems.
Because of this, we can use comparisons between the predicted 
    and observed isolated binary populations to infer constraints on
    models of binary evolution and learn about the underlying astrophysical processes.

As proof of concept, comparable methods have been applied to interpret the population of sources identified by
ground-based gravitational wave detectors.
The LVK has published approximately ninety gravitational wave observations
    in the Gravitational Wave Transient Catalog (GWTC)
    \citep{GWTC-1, GWTC-2, GWTC-3, GWTC-2p1}.
The population of these observations informs us about the underlying population
    of merging compact binaries
    \citepLVKRP{}.
These merger events are expected to form from specific formation channels,
    such as the isolated evolution of massive stellar binaries
    \citepLVKIsolated{} ,
    and various dynamic formation scenarios
    \citepLVKDynamic{};  see \citet{MANDEL20221} for a detailed review of formation channels.
By comparing gravitational wave observations to the results of
    binary evolution population synthesis models,
    we can use Bayesian inference to 
    learn about the underlying astrophysics
    \citep{COSMICZevin2021,Stevenson2017,st_inference_interp}.
Within the context of these ground-based instruments, previous studies have used synthetic populations of sources within
an end-to-end calculation to assess how well these calculations can constrain source model parameters, a process we
denote for brevity as ``injection recovery''
    \citep[see, e.g.,][]{Barrett2017,Wysocki2019,COSMICZevin2021} and references therein.

In this paper, we adapt the Bayesian hierarchical inference methods used
    to infer constraints on the formation of massive compact binaries
    from the LVK observations 
    \citep[e.g.][]{st_inference_interp} to study the formation of
    LISA Galactic DWD binaries.
We choose to focus on common envelope evolution in this paper
    as it is not well constrained by existing studies, and
    may have a powerful impact on the population of 
    Galactic DWD binaries observable by LISA
    \citep{Korol2022CE}.
Furthermore,
    we demonstrate the recovery of an injected astrophysical assumption
    about the common envelope efficiency parameter
    from the population of Galactic DWD gravitational wave sources
    with resolved chirp mass estimates.

This paper is organized such that each section describes the methods
    pertaining to a stage of our investigation, as well as 
    results pertaining to that stage.
In section \ref{sec:cosmic}, we describe our methods for simulating
    the evolution of DWD binaries with the COSMIC binary population synthesis code.
    We explore the impact of common envelope ejection efficiency
    on different formation scenarios
    for DWD binaries.
In section \ref{sec:galaxy}, we explore the star formation environments in a synthetic galaxy,
    and identify the subpopulation of DWD binaries observable by LISA
    for our selection of formation models.
In section \ref{sec:likelihood}, we describe our Bayesian hierarchical
    inference methods and their impact on populations of those simulated binaries
    which could be resolved above the stochastic gravitational wave background of 
    the galaxy.
In Section \ref{sec:kl}, we adopt an alternative approach investigating the
    usefulness of the KL divergence statistic on the populations of resolved binaries,
    in recovering an injected formation model.
Finally, in section \ref{sec:discussion}, we summarize the impact of our results
    and suggest how it may inform ongoing and future studies.
 \section{Simulating a population of Galactic DWD binaries}
\label{sec:cosmic}
There are many uncertainties remaining in the evolution of stellar
    binaries, such as the initial binary fraction \citep{Thiele2023}
    and various forms of mass transfer initiated by Roche overflow interactions.
One such form of dynamically unstable mass transfer is common envelope evolution,
    where the atmosphere of a donor star expands beyond its Roche lobe such that both stars become immersed in
    a common envelope.
The most accurate method for characterizing the efficiency of common envelope
    evolution (in ejecting orbital energy and angular momentum from the system)
    is currently debated by various groups 
    \citep[e.g.][]{IvanovaRemnant2011,IvanovaEnthalpy2011,Klencki2021,Wilson2022,
    Hirai2022,DiStefano_2023,Tang2024}.
\correction{
If the common envelope prescription employed in the simulation of DWD
    binaries fails to describe the onset and outcomes of common envelope evolution,
    this may account for a mismatch between the properties
    of observed and simulated populations.
}

It has been proposed that gravitational wave observations
    are an effective way to probe models of binary evolution
    and constrain these uncertain processes
    \citep{Barrett2017,Belczynski2020-EvolutionaryRoads,Thiele2023,
    st_inference_interp}.
The first step in predicting a population of observable DWD binaries
    in the LISA band is to use a population synthesis 
    code to simulate a population of
    DWD binaries for an assumed set of astrophysical assumptions 
    and initial binary parameter distributions.
We perform these simulations for different kinds of DWD binaries
    with different Zero Age Main Sequence (ZAMS) metallicity assumptions,
    and considering the evolution of a Galactic population from
    the dawn of cosmic stellar formation to present day.

\begin{subsection}{Binary Population Synthesis with COSMIC}
We use COSMIC \citep{Breivik2020, COSMIC_code}
    to simulate populations of DWD binaries. 
COSMIC is an open-source 
    rapid binary population synthesis suite that is based on the single star
    evolution `Hurley' fits \citep{Hurley2000} to the 
    `Pols 1990' \citep{Pols1998} stellar evolution tracks and binary evolution 
    algorithm of \citep{Hurley2002}.

The primary difference between COSMIC and other rapid 
    population synthesis tools like SeBa, \citep{PortegiesZwart1996}
    MOBSE \citep{Giacobbo2018}, 
    \texttt{StarTrack} \citep{Ruiter2010}
    or COMPAS \citep{Riley2022},
    is the use of the $match$ statistic, an iterative population 
    sampling criteria inspired by matched filtering techniques.
Instead of implementing an adaptive importance sampling technique 
    like STROOPWAFEL's implementation in COMPAS 
    \citep{2019MNRAS.490.5228B}, the $match$ statistic allows us to continuously
    draw samples until the shape of density histograms
    of binary parameters (like mass and orbital period) in the selected population
    become numerically stable.
For a detailed discussion of the $match$, see \citet{Breivik2020}.

COSMIC classifies WD remnants by their chemical composition;
    Helium dominated WDs are labeled ``He'', while
    Carbon/Oxygen dominated WDs are labeled ``CO'',
    and Oxygen/Neon dominated WDs are labeled ``ONe''.
Following \citet{Thiele2023},
    we separately simulate the HeHe, COHe, and COCO combinations,
    and collectively simulate all systems with an ONe WD (labeled ONeX).
In this work, we refer to these classifications (HeHe, COHe, COCO, and ONeX)
    as ``kinds'' of DWD systems.
This is primarily so that the $match$ criteria can accurately 
    capture the mass distributions of each WD binary class.
In the case of ONeX binaries, however, the number of these systems 
formed per unit solar mass of stars formed is significantly lower.
Hence, we combine all companion WD classes for these binaries.
Also following \citet{Thiele2023},
    we simulate populations of each type of DWD in 15 metallicities,
    distributed uniformly in the log-space of the interval $[10^{-4}, 0.03]$,
    consistent with the limits of the Hurley fits to the Pols stellar evolution tracks
    \citep{Pols1998,Hurley2000,Hurley2002}.

The primary mass of each binary is sampled from a Kroupa 2001 IMF \citep{Kroupa2003},
    while the secondary is drawn from a uniform distribution in mass ratio
    ($q \in [0.01,1.0]$).
COSMIC assumes a metallicity-dependent close binary fraction
\correction{
    (i.e. fraction of stellar binaries with initial orbital periods less than six days)
}
    consistent with \citet{Moe19} that decreases from $50\%$ at Z=0.0001 to $15\%$ at Z=0.02.
The orbital periods are distributed log-normally with 
    increased numbers of systems below $10,000\,\rm{days}$ 
    at sub-solar metallicities. We further assume that initial 
    orbital eccentricities are uniformly distributed between 0 and 1.

\begin{table}[!ht]
\centering
\begin{tabular}{|c|c|c|}
\hline
Model ID & sampling & $\alphaCE$ \\
\hline
CEb50-CEb70 & linear & $[0.5, 1.5]$ \\
\hline
CEb71-CEb81 & logarithmic & $[0.1, 10.]$ \\
\hline
\end{tabular}
\caption{\label{tab:CEmodels}
\textbf{Common envelope assumptions for COSMIC models:}
The label and corresponding value of $\alphaCE$ for each model included in this study.
For reference, CEb60 is our model for $\alphaCE = 1.0$.
}
\end{table}

COSMIC contains several updated prescriptions that treat binary-star
    interactions. For a detailed discussion of these updates, see
    \citet{Breivik2020}.
We broadly apply the default assumptions for binary interactions 
    as indicated in the documentation for V.3.4.10 of COSMIC.

As noted by \citet{Thiele2023}, assumptions for common envelope
    evolution can play a significant role in the the size and characteristics of
    the DWD population.
In this work, we use an energy conservation argument 
    \citep[e.g.][]{Paczynski1976} that balances the binding energy of the envelope
    with the change in orbital energy required to eject the envelope.
In this case, we use the common envelope ejection efficiency, $\alphaCE$
    to characterize this process,
    and allow it to vary broadly in order to avoid the exclusion of models 
    which account for phase transitions and other astrophysical effects.
The case where $\alphaCE=1$ represents a perfect conversion of the orbital 
    energy reservoir to be used in unbinding the envelope.
We assume that the binding energy of the envelope, often characterized 
    by a $\lambda$ parameter, is modeled based on the envelope structure of 
    the donor \citep[as described in Appendix A of][]{Claeys2014}.
We note that $\alphaCE$ and $\lambda$ parameters enter into our common envelope 
    ejection model as
\begin{equation}
    \frac{E_{\rm{bind,i}}}{\lambda} = \alphaCE (E_{\rm{orb,i}} - E_{\rm{orb,f}}).
\end{equation}
Thus, by keeping our binding energy model assumptions fixed, changes 
    in $\alphaCE$ may be equivalently capturing different descriptions for the binding 
    energy parameter $\lambda$.
Indeed, the case of $\alphaCE>1$ could be due to additional energy sources 
    or due to donors with less 
    bound envelopes than our model assumptions \citep{Yamaguchi2024a,Yamaguchi2024b}.

For common envelope evolution,
    the binding energy 
    in the implementation of COSMIC we use here
    is estimated as in Appendix A of \citet{Claeys2014}.
For the common envelope ejection efficiency ($\alphaCE$),
    we assume a small family of models with values ranging 
    logarithmically between $0.1$ and $10$, 
    and a more finely sampled family of models with values
    ranging linearly between $0.5$ and $1.5$
    (see Table \ref{tab:CEmodels}). 
We chose these values to bracket the wide uncertainty associated with
    common envelope ejection, though we note that multiple studies have found
    ejection efficiencies near $\alphaCE\sim0.3$ for WD binaries \citep[e.g.][]{Zorotovic2010,Scherbak2023} 
    or a potential scaling of $\alphaCE$ with the mass of the donor star \citep[e.g.][]{DeMarco2011}.
Our results include 480 COSMIC simulations of DWD populations
    (for 32 choices of $\alphaCE{}$ and 15 choices of ZAMS metallicity).
Both CEb60 and CEb76 have $\alphaCE{} = 1.0$,
    where CEb60 is considered our fiducial (or injected) population
    and CEb76 was used to examine the impact of random number generation
    within COSMIC on statistical inference.

\end{subsection}
\begin{subsection}{Gravitational wave evolution and delay time}
All close DWD binaries shrink due to gravitational wave
    radiation from the time of DWD formation to present day.
The LISA binaries will also shrink during the lifetime of the LISA mission.
Orbital evolution due to gravitational wave radiation
    in a circular binary
    is carried out using the definitions from
    \citet{Peters1964}:
\begin{equation}\label{eq:gwsep}
    a_f = (a_i^4 - 4 \beta t_{\mathrm{delay}})^{1/4}
\end{equation}
\begin{equation}\label{eq:gwtimescale}
    \beta = \frac{64G^3}{5c^5} m_1m_2(m_1 + m_2),
\end{equation}
where $a_i$ and $a_f$ are the initial and final separation
    of the binary, and $m_1$ and $m_2$ are the mass of each object.

Ultimately, the quantities which describe potential gravitational wave observations
    are masses and orbital periods.
These are related to the star forming conditions via a delay time:
    the time between the zero age main sequence formation of 
    the binary and a gravitational wave observation.
To track this evolution properly, we characterize each binary by
    the separation and mass of each remnant at the time of
    the formation of the white dwarf.
As DWD orbital frequencies continue to evolve over time
    according to gravitational wave radiation, 
    we calculate this ``delay time'' as the sum of the time required to form the compact binary
    ($t_{\mathrm{form}}$)
    and the time required to evolve the system to a final orbital frequency
    at present day
    ($t_{\mathrm{evol}}$).
Therefore, we draw several quantities from the output of COSMIC simulations:
    the separation of the binary at DWD formation,
    the mass of each remnant,
    and the time of compact binary formation.
Because Roche lobe overflow 
    is assumed to fully circularize close DWDs, 
    which are the products of either stable mass transfer or common envelope, 
    all DWD binaries analyzed in this work are assumed to form as circular binaries.
We also interpret COSMIC outputs to track the prevalence of different chemical compositions
    in the underlying white dwarf populations.

\begin{figure*}
\centering
\includegraphics[width=\textwidth]{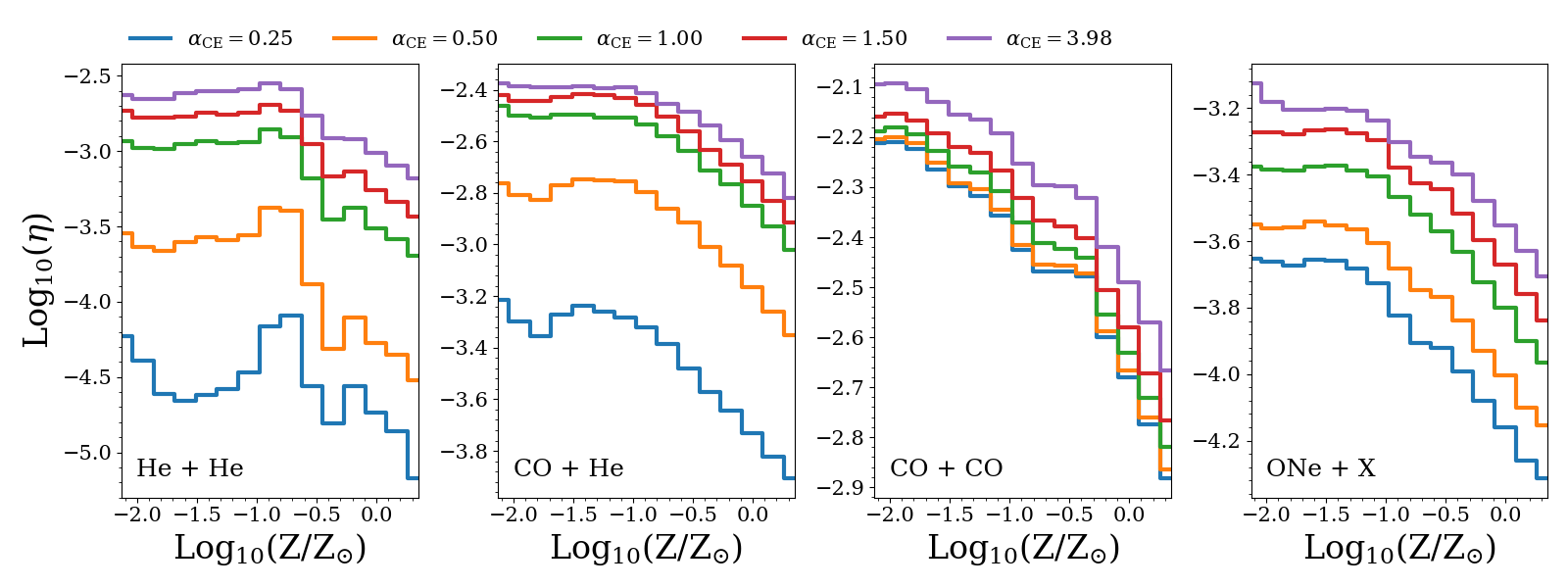}
\caption{\label{fig:form_eff}
    \textbf{The DWD formation efficiency as a function of metallicity:}
    Each curve represents the number of DWDs which form per solar mass of star formation,
        for a given type of DWD and for a given metallicity.
    The color of each curve indicates the $\alphaCE$ assumption
        for a batch of COSMIC runs.
}
\end{figure*}

Gravitational wave radiation can drive some binaries to transfer mass
    or contact before present day.
\correction{
These systems can exchange mass or merge 
    during the DWD stage of a binary's life.
}
For these systems, the ``delay time'' is the time between
    star formation and contact
    (rather than between star formation and present day).
The fate of such a system can be hard to determine given our current state
    of knowledge, but may depend on many things including the mass ratio
    of the binary \citep{Marsh2004,Gokhale2007,Sepinsky2010,Kremer2015,Kremer2017}.
While these binaries may be observable by LISA,
    they will be identified separately from the detached binary population
    \citep{Breivik2018WD},
    and we consider them beyond the scope of this work.
\correction{
Our assumptions about how close a DWD binary must come
    (i.e. the critical separation)
}
    before mass transfer may occur are consistent with \citet{Thiele2023}.

On the other hand, some binaries may remain too wide
    to fall inside of LISA's sensitivity.
In this work, we consider DWD binaries with frequencies
    that satisfy $f_{\mathrm{GW}} \in [10^{-4}, 1.0] \mathrm{Hz}$ at the present day,
    consistent with LISA's nominal sensitivity requirements \citep{LISASensitivity}.

Section \ref{sec:detection} explores which binaries will have resolved 
    parameter estimates,
    and which will contribute to the Galactic foreground.
\end{subsection}

\begin{subsection}{Results of changing $\alphaCE$ on binary formation pathways}

For each COSMIC simulation,
    we obtain both the number of DWDs of a given kind 
    ($N_{\mathrm{DWD,sim}}(\metallicity,\mathrm{kind})$)
    and the total mass of all binaries sampled
    ($M_{\mathrm{ZAMS,sim}}(\metallicity,\mathrm{kind})$)
    to produce the DWDs in our simulated population.
This allows us to estimate the formation efficiency of DWDs per unit solar mass of star formation, 
\begin{equation} \label{eq:form_eff}
\eta_{\mathrm{form}}(\metallicity, \mathrm{kind}) = 
    N_{\mathrm{DWD,sim}}(\metallicity,\mathrm{kind}) /
    M_{\mathrm{ZAMS,sim}}(\metallicity,\mathrm{kind}),
\end{equation}
    for each kind of DWD binary (HeHe, COHe, COCO, ONeX),
    at each COSMIC metallicity.

Figure \ref{fig:form_eff} exhibits the formation efficiency of DWD binaries for
    a few choices of $\alphaCE$, selected to encompass the range of all simulated alphas.
We note that as $\alphaCE$ increases, $\eta_{\mathrm{form}}$ increases as well;
    broadly, this is because fewer systems undergo pre-DWD stellar mergers when 
    common envelope ejection is more efficient.
We point the reader to Section 5 of
    \citet{Thiele2023} for a related discussion of formation efficiency trends.
We also note that this scaling is not applied uniformly across all kinds 
    of DWD binaries and metallicities
    (note the varied y-axis ranges in Figure \ref{fig:form_eff}).
The common envelope ejection efficiency has a stronger impact on binaries 
    containing He WDs because their progenitors have smaller radii relative to 
    CO or ONe WD progenitors. 
This means that the common envelopes which produce He WDs occur in orbits with 
    initially shorter periods and thus a smaller orbital energy well to be used 
    in unbinding the envelope from the system. 
As the common envelope ejection efficiency decreases, we find a significant 
    reduction in the number of DWDs containing He WD due to mergers caused by \
    unsuccessful envelope ejection.
\correction{
This motivates a search for further quantitative (statistical)
    dependence on $\alphaCE$ through
    i) the total number of binaries,
    ii) the relative contribution of different formation pathways
    to the overall population, and
    iii) the shape of the density in observable quantities.
}

\end{subsection}
 \section{A synthetic galaxy of LISA sources}
\label{sec:galaxy}

In order to simulate a LISA population of DWD binaries,
\correction{
    we need to assume a star formation history
    by sampling from star formation in a synthetic galaxy.
}
We fill this synthetic galaxy with binaries simulated by COSMIC
    to construct a Galactic DWD population.
Following this, we classify binaries
    based on whether they have time to form and
    remain in the LISA band until present day.
Finally, the binaries resolved above the confusion limit
    are identified from the sample population
    and their uncertainties are estimated using a Fisher matrix
    method with the ldasoft code \citep{ldasoft}.

\begin{subsection}{Sampling star formation with FIRE-2 m12i}
\begin{figure}
\centering
\includegraphics[width=3.375 in]{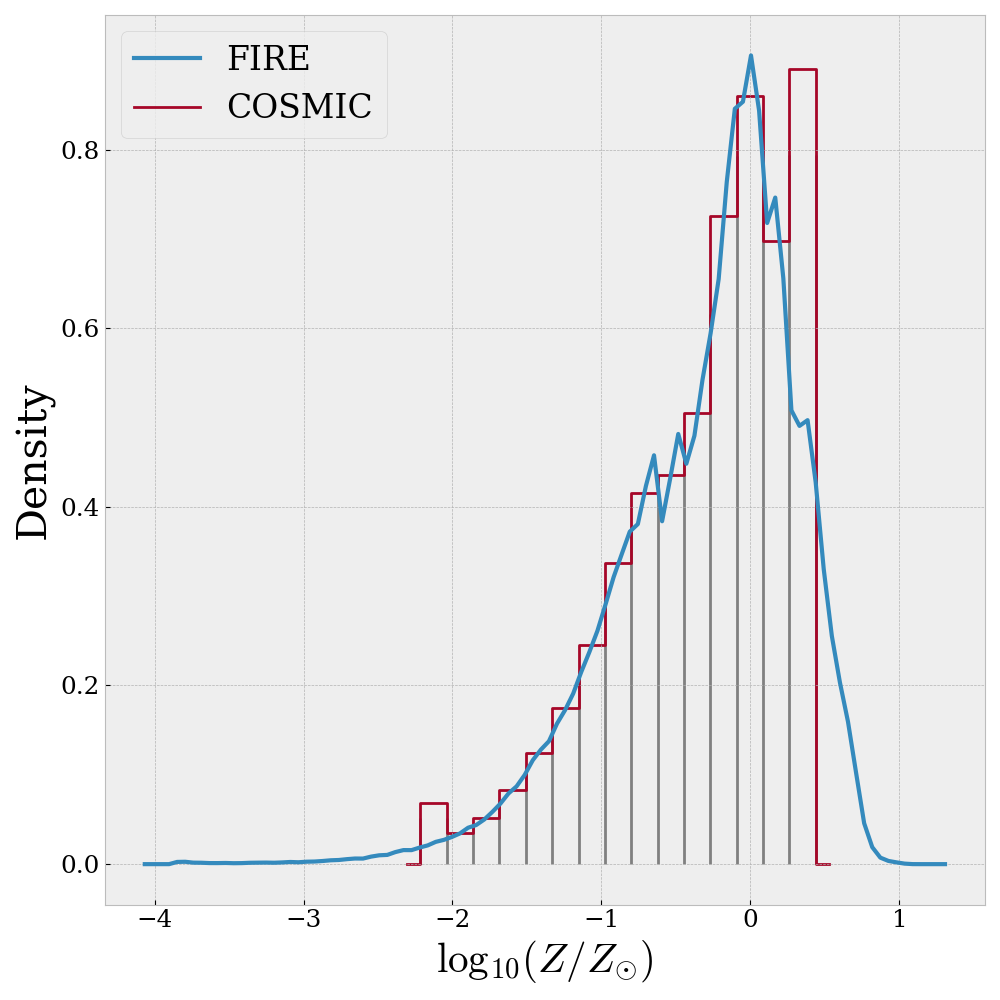}
\caption{\label{fig:metmet}
    \textbf{Sampling from discrete metallicity bins:}
The DWD populations are simulated on a fixed grid of metallicities, 
        and the FIRE-2 star particles have a 
        metallicity distribution which varies continuously.
    Therefore, each star particle must be sorted into the closest COSMIC
        metallicity bin.
    This figure features histograms for both the FIRE star particle metallicities
        (100 bins) and the 15 COSMIC metallicity bins.
    This figure demonstrates an example for the CEb60 model 
        ($\alphaCE = 1.0$),
        in order to broadly contrast the continuous and discrete
        metallicity distributions.
}
\end{figure}
In this work, we follow \citet{Thiele2023}, and sample
    star formation from a Milky-Way-like synthetic galaxy,
    \textbf{m12i} \citep{Wetzel2016}, using the Feedback In Realistic Environment 2
    (FIRE-2) simulations
    \citep{Hopkins2014,Hopkins2015,Hopkins2018}.
This simulated galaxy was generated using the GIZMO code \citep{Hopkins2015}
    to solve hydro-dynamical equations and evolve Galactic morphology
    and star formation incorporating stellar feedback and metallicity evolution across cosmic time.
The Latte suite of the FIRE-2 simulations
    follow the evolution of Milky-Way-like galaxies (such as \textbf{m12i})
    \citep{Wetzel2016, Garrison-Kimmel2017, Sanderson2020}.

The star formation history of the \textbf{m12i} synthetic galaxy is represented
    by discrete packets of star formation, henceforth referred to as ``star particles''.
Each star particle represents $7070 \msun$ of star formation.
The star particle samples from the database for \textbf{m12i} include
    local satellite galaxies.
Consistent with \citet{Lamberts2019}, we distinguish our central galaxy by defining
    as selection within $300 \mathrm{kpc}$ of the center of the galaxy.
In practice there is no need to enforce this selection,
    as there are no DWD systems outside of this region which meet
    the LISA signal-to-noise-ratio cut we use to define the resolved population
    (see Section \ref{sec:detection}).

We use the time of star formation,
    the metallicity, and position of each star particle 
    to assign formation times and positions for binaries simulated by COSMIC.
We sort these into metallicity bins,
    corresponding to the closest discrete COSMIC simulation metallicity
    (see Figure \ref{fig:metmet}).
We estimate the number of binaries per star particle
    for each metallicity bin, according to our formation efficiency
    (Eq. \ref{eq:form_eff}):
    $N_{\metallicity, \mathrm{kind}} = \eta_{\mathrm{form}}(\metallicity,\mathrm{kind}) M_{*}$,
    where $M_{*}$ is the mass of a star particle ($7070 \msun$).
For each star particle, the number of samples drawn from the corresponding
    COSMIC simulation starts at the integer portion of
    $N_{\metallicity, \mathrm{kind}}$, with an additional sample being
    drawn a fraction of the time equal to the decimal portion.

Previous work \citep{Thiele2023} used the Ananke framework \citep{Sanderson2020}
    to populate the \textbf{m12i} galaxy with DWD systems,
    offset to the center of each star particle by a random distance.
In this work, we use the output from the Latte simulations directly,
    and note that for our fiducial model there are close to (or less than)
    one LISA-band DWD binary for each star particle.
We therefore populate the galaxy with DWD binaries
    at the center of each star particle.

Within COSMIC, solar metallicity ($\zsun=0.02$)
    is consistent with \citet{Tout1996},
    to which the the radii and luminosity power-laws from BSE are calibrated.
This Tout et al. (1996) metallicity assumption is a poor estimator of
    iron abundances, as elemental abundances are estimated from meteorites
    \citep{Anders1989}.
Meanwhile, in FIRE-2, solar metallicity is assumed to be $\zsun=0.01342$,
    calibrated to \citet{Asplund2009}.
For this reason, absolute metallicity 
\correction{
 (i.e. the base 10 logarithm of the fraction of elements with an atomic number of 3 or higher)
 of \textbf{m12i} star particles 
}
is used to inform
    this metallicity-dependent sampling process,
    as using different solar metallicity assumptions for
    FIRE-2 and COSMIC star formation can otherwise introduce a bias. 
\end{subsection}

\begin{subsection}{Confusion Limited Sources}
\label{sec:detection}
We use LEGWORK \citep{LEGWORK} to calculate the strain amplitude of each source
    in the second orbital frequency harmonic for the LISA instrument,
    assuming a LISA observation time of 4 years
    \citep[see Section 3 of][]{Thiele2023}.
We then use ldasoft \citep{ldasoft}
    to estimate the Signal-to-Noise-Ratio (SNR)
    of each source above the Galactic foreground.
With ldasoft, we analyze the LISA time domain interferometry
    across the AE channels,
    and estimate the confusion noise using a running median
    across frequency bins with a resolution determined
    by $1/\tobs = 1/4 \mathrm{yr}^{-1} \approx 8 \times 10^{-9} \mathrm{Hz}$.
As well-resolved sources can be removed from the confusion of unresolved sources
    which make up the Galactic foreground,
    we subtract the brightest sources,
    recalculate the impact of the Galactic foreground on the population of resolved DWDs,
    and generate updated SNRs for each source.
Finally, we also use ldasoft to estimate the inverse covariance matrix
    for high SNR ($\geq 7$) sources using the Laplace approximation
    \citep{flournoy1991statistical, Cutler1994FisherGW}.
For the rest of this work, 
    these sources are called the ``resolved'' population.

This approximation
    fits a Gaussian distribution to the posterior probability of each source:
\begin{equation}\label{eq:laplace-approx}
    \pprob(\BinaryParameters|\FormationParameters) \approx 
    \pprob(\BinaryParameters|\FormationParameters)|_{\BinaryParameters=\BinaryParameters_{*}}
        \exp \Big(
        -\frac{1}{2}(\BinaryParameters - \BinaryParameters_{*})^{T}
        \Gamma^{-1}
        (\BinaryParameters - \BinaryParameters_{*})
        \Big),
\end{equation}
where Fisher information $\Gamma$ is given by
\begin{equation}\label{eq:fisher}
\Gamma^{-1} = - \nabla_{\BinaryParameters}\nabla_{\BinaryParameters} 
    \mathrm{ln}(\pprob(\BinaryParameters | \FormationParameters))\Big|_{
    \BinaryParameters = \BinaryParameters_{*}},
\end{equation}
and where $\BinaryParameters$ represents an arbitrary guess at the true
    parameters of a source, $\BinaryParameters_*$,
    and $\FormationParameters$ defines our other assumptions.

We use this Gaussian approximation to assume the form of our
    likelihood function $\gwL(\BinaryParameters)$,
    where our binary parameters are
    gravitational wave frequency ($f_\mathrm{GW}$),
    the time derivative of that frequency ($\dot{f}_{\mathrm{GW}}$),
    and strain amplitude ($h$).
Following this, we use the Jacobian transformation to change coordinates
    to $f_{\mathrm{GW}}$, chirp mass ($\mc$), and distance ($\lumdist$).
We then marginalize over distance to construct a likelihood 
    for each source, $\gwL(\fgw, \mc)$.
\correction{
In gravitational wave Parameter Estimation (PE),
    Fisher estimates for parameter uncertainties can encounter numerical challenges,
    as well as bias (when astrophysical prior information is not fully considered).
These and other limitations of Fisher estimates
    are summarized well by \citep{Vallisneri2008}.
}
However,
    as the number of sources can vary widely (up to tens of millions of sources),
    estimating a detailed likelihood approximation more detailed than a Fisher estimate
    for each source
is beyond the scope of this work.

A reference prior is typically assumed for single-event parameter inference
    for the parameters of gravitational wave observations of massive compact binaries
    \citep{Veitch2015LALInference, Bilby2019, CallisterPrior2021}.
As this reference prior would factor out of our parameter estimates for 
    synthetic DWD binaries, we do not need to adopt one.

\end{subsection}

\begin{subsection}{Results of changing $\alphaCE{}$ on Galactic binary populations}
\label{sec:1d}

\begin{figure}
\centering
\includegraphics[width=3.375 in]{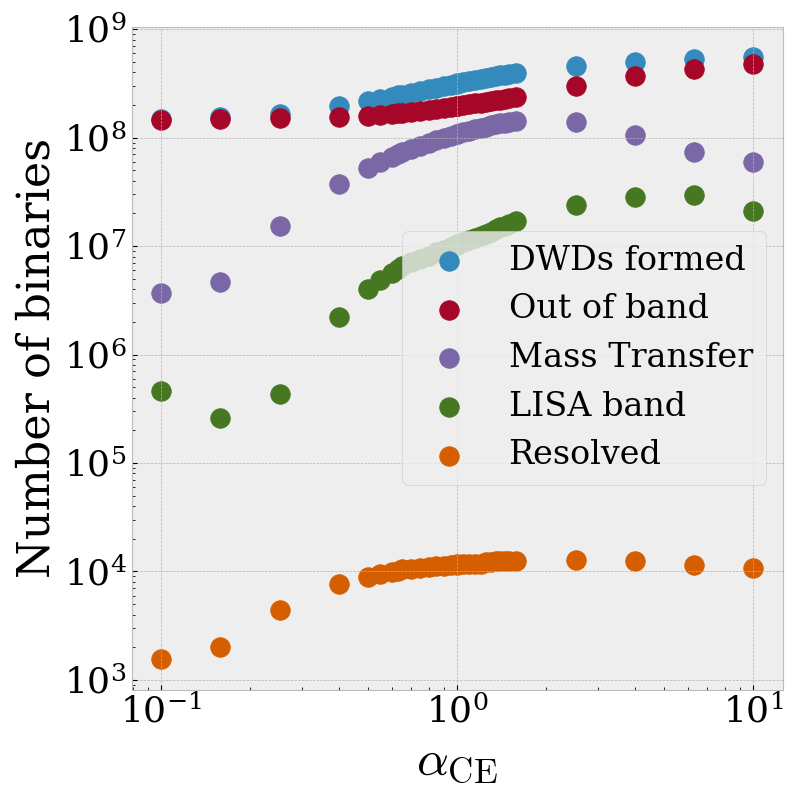}
\caption{\label{fig:m12i_Numbers}
    \textbf{Binary classifications in $\alphaCE{} $:}
Classifications of a simulated Galactic DWD population at
    different choices of $\alphaCE{}$, for the \textbf{m12i} galaxy.
}
\end{figure}

After sampling a Galactic DWD population from
    star formation assumptions for the \textbf{m12i} galaxy,
    we identify a Galactic binary population for each choice of 
    our formation assumptions.
We have explored 32 sets of COSMIC simulations with different choices
    for $\alphaCE$.
The assumed common envelope for each simulation can be found in Table \ref{tab:CEmodels}.

\begin{figure*}
\centering
\includegraphics[width=2.250in]{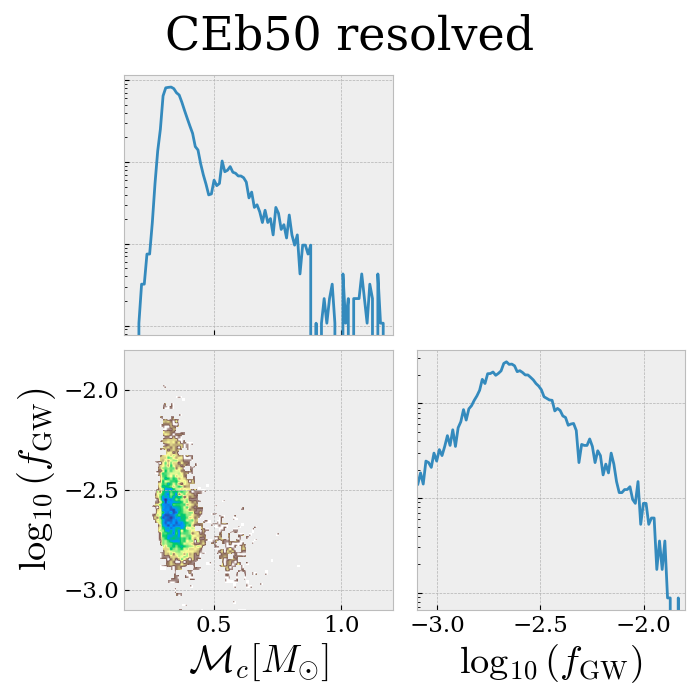}
\includegraphics[width=2.250in]{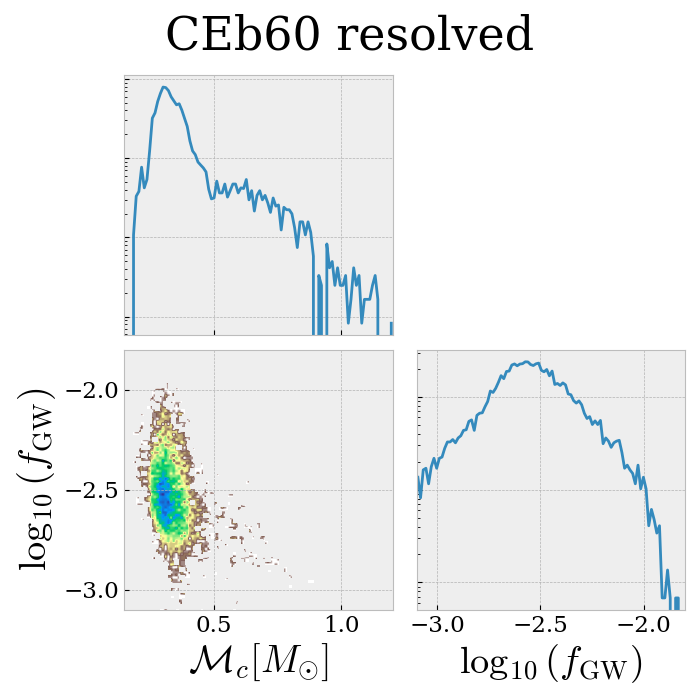}
\includegraphics[width=2.250in]{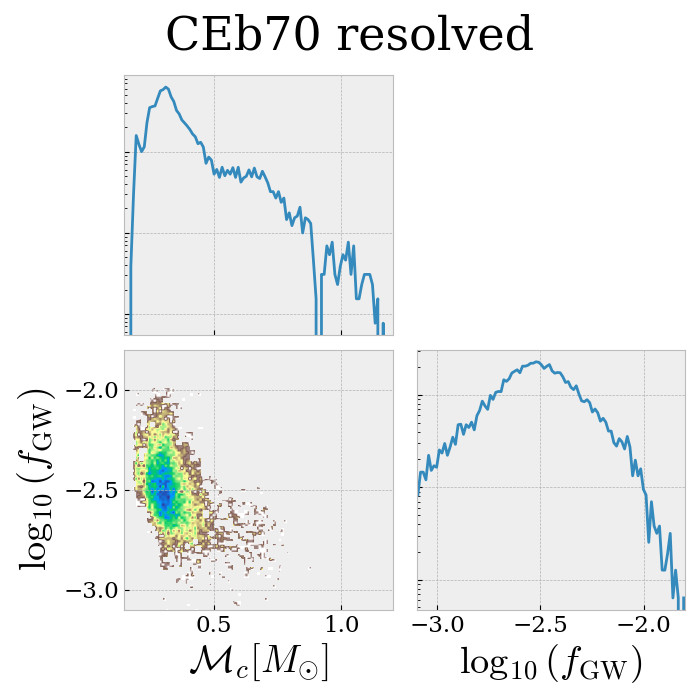} \\
\includegraphics[width=2.250in]{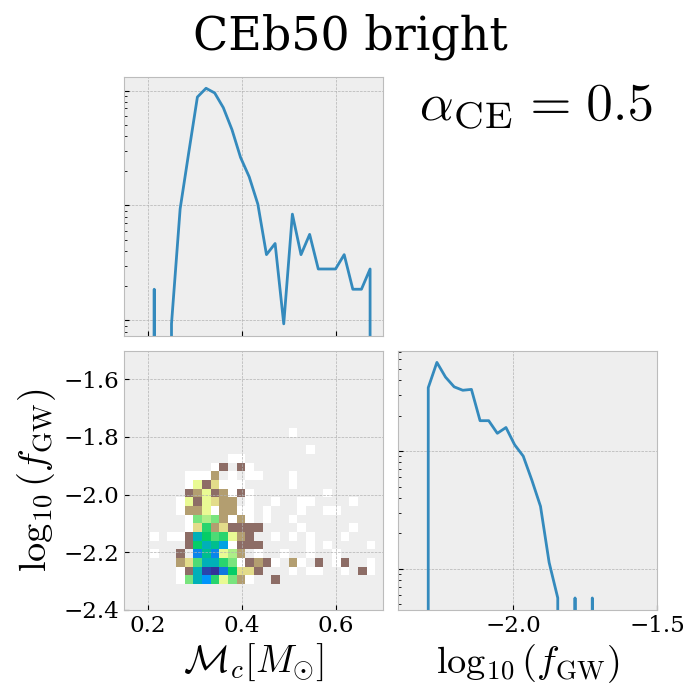}
\includegraphics[width=2.250in]{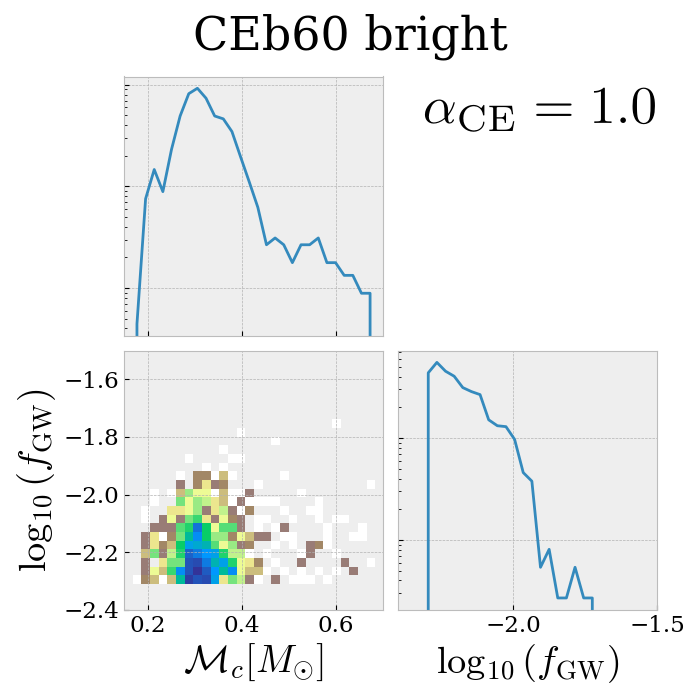}
\includegraphics[width=2.250in]{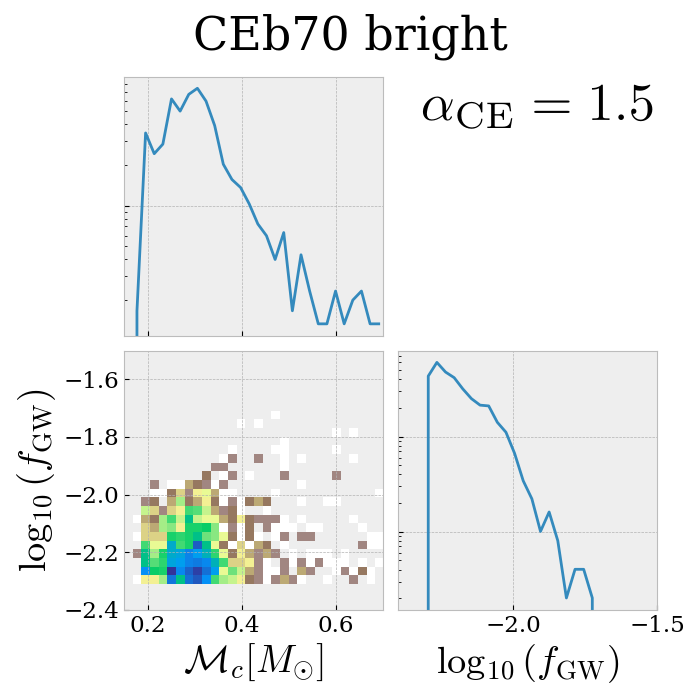}
\caption{\label{fig:corner_intrinsic}
\textbf{Intrinsic parameters of the resolved population:}
    Each corner plot exhibits the one- and two-dimensional
        distribution for the resolved (top) or LISA-bright (bottom) population
        of Galactic DWD binaries, in $\mathrm{log}_{10}$ of
        gravitational wave frequency and chirp mass.
    The resolved (top) plots identify sources for which $\mathrm{SNR} > 7$.
    The LISA-bright (bottom) plots identify sources for which
        $\mathrm{SNR} > 20$ and $\fgw > 5 \mathrm{mHz}$.
    From left to right: CEb50 ($\alphaCE = 0.5$),
        CEb60 ($\alphaCE = 1.0$), and CEb70 ($\alphaCE = 1.5$) models.
    Each diagonal plot contains one-dimensional histograms for the intrinsic parameters,
        while the off-diagonal shows a two-dimensional histogram.
    Limits and bins are consistent across each row.
    Chirp mass is truncated at $\mc \leq 0.7 \msun$.
}
\end{figure*}

The impact of common envelope efficiency on 
    the number and classification of sources in the synthetic Galactic DWD population
    is demonstrated in Figure \ref{fig:m12i_Numbers}.
This figure accounts for the close binaries within the LISA gravitational wave
    frequency band, those wide binaries which are outside of that band,
    and binaries which come close enough together to initiate 
    mass transfer during the DWD phase.
The number of LISA band sources can vary by orders of magnitude
    with different $\alphaCE{}$ assumptions. 
This is a direct consequence of models with less efficient common envelope ejection 
    leading to more stellar mergers that occur 
    before a DWD can form (c.f. Figure~\ref{fig:form_eff}).

We define the ``LISA-bright'' population as the sources with $\mathrm{SNR} > 20$,
    and $f_{\mathrm{GW}} > 0.005$.
This population will be used for inference,
    to ensure strong PE results.
After also applying the frequency cut,
    this SNR cut results in 1237 LISA-bright sources
    for our fiducial model (CEb60),
    and does not significantly reduce the number of LISA-bright sources
    compared to an $\mathrm{SNR}$ cut of 7,
    which would result in 1255 resolved sources.
Figure \ref{fig:corner_intrinsic} shows how the density of binaries
    (in their intrinsic parameters $\fgw$ and $\mc$)
    changes with these cuts.
It is worthy of note that the number of resolved and LISA-bright 
    sources varies by fewer orders
    of magnitude than the total number of LISA band systems.
This is because an increase in the number of
    LISA band sources causes an increase in confusion noise
    in the Galactic foreground, affecting sensitivity to individual sources.

The impact of common envelope ejection efficiency on the Galactic binary population
    extends further than just the number of predicted LISA-bright sources.
Qualitatively, we observe subtle differences between the selected populations
    (see Figure \ref{fig:corner_intrinsic}).
For example, the low $\alphaCE{} $ model lacks sources near the lower end of the mass range,
    and tends toward lower frequencies.
This is because the lowest mass He WD progenitors interact in significantly 
    closer binary configurations than their more massive counterparts. 
The orbital energy reservoir that is used in the common envelope ejection is thus much smaller. 
When combined with a lower common envelope ejection efficiency, these systems are more likely 
    to merge prior to the formation of a DWD and thus not become LISA sources. 
For $\alphaCE{} $ greater than unity, we notice a distinct mode in the lower mass range 
    that emerges due to these systems no longer merging before DWD formation.
In the rest of this paper, we examine if these features are statistically robust
    enough to place common envelope model constraints using the
    LISA Galactic DWD population.

\end{subsection}
 \section{How to constrain binary evolution models with $\alphaCE$}
\label{sec:likelihood}
In the previous section, we identified our methodology for
    predicting a population of detached DWD binaries observable
    by LISA with resolved chirp mass estimates.
By changing the assumptions of our binary evolution model,
    we can measure the impact of these assumptions on 
    predicted populations.
Past work constraining populations of stellar mass black hole binaries
    using LVK observations has either made use of 
    phenomenological population models
    or of likelihood models for individual observations
    constructed with the parameter estimation samples in the GWTC releases
    \citep{GWTC-1, GWTC-2, GWTC-2p1, GWTC-3, GWTC-2p1-Zenodo, GWTC-3-Zenodo,
    nal-chieff-paper, nal-methods-paper}.
In our own work, we have taken the latter approach
    \citep{st_inference_interp}.
By integrating over the population of mergers predicted by binary population
    synthesis, past work was able to demonstrate a method for
    constraining the underlying assumptions in those population models.

We aim to demonstrate that in principle
    we can use the same approach to constrain LISA populations.
In this work, we inject a population of LISA-bright DWD Galactic binaries
    with a known value of common envelope ejection efficiency ($\alphaCE$),
    and attempt to recover $\alphaCE$
    by comparing the injected population to test populations.
We define the ``LISA-bright DWD'' population of a simulated galaxy
    (discussed in Section \ref{sec:cosmic})
    to be those detached DWD binaries for which
    $\mathrm{SNR} > 20$ and $\fgw > 5$ mHz,
    with covariance estimates from Fisher matrix approximations
    (described in Section \ref{sec:detection}).

Our approach relies on statistically characterizing how
    closely any test population resembles the injected population,
    both in count and properties.
We estimate the likelihood of the injected data,
    given our population model.   
In this section, we detail the methods used to construct
    and evaluate the Inhomogeneous Poisson Point Process likelihood 
    likelihood to draw constraints on injected astrophysics.
We also discuss the role of marginalization
    over various realizations of a test population.

Finally, we demonstrate the effectiveness
    of this likelihood model through the example of
    a one-dimensional injection recovery in $\alphaCE$.

\begin{subsection}{The Inhomogeneous Poisson Point Process likelihood}

In general, when we run an experiment and measure some data, $d$,
    we can use Bayes' theorem to evaluate the likelihood of some
    set of hypotheses, $\mathcal{H}$:
\begin{equation}
\label{eq:bayes}
\pprob(\mathcal{H}|d) = \frac{\pprob(d|\mathcal{H}) \pprob(\mathcal{H})}{\pprob(d)}.
\end{equation}
where $\pprob(\mathcal{H}|d)$ is the posterior probability of our hypotheses,
    $\pprob(d|\mathcal{H})$ is the likelihood of our data given our set of hypotheses,
    $\pprob(\mathcal{H})$ is the probabilistic representation of our prior knowledge about
    $\mathcal{H}$, and $\pprob(d)$ is a marginalization factor to keep our probability
    distribution normalized.
That marginalization factor can be calculated over all possible
    hypotheses:
\begin{equation}
\pprob(d) = \int d\mathcal{H} \pprob(d|\mathcal{H}) \pprob(\mathcal{H}).
\end{equation}
Marginalization also allows us to reduce the dimensionality
    of our set of hypotheses:
Suppose that our set of hypotheses depend on two parameters,
    $\lambda_1$ and $\lambda_2$.
We can also marginalize over
    $\lambda_1$ to find our likelihood in terms of $\lambda_2$:
\begin{equation}
\pprob(d | \lambda_2) = \int_{\lambda_1} \pprob(d | \lambda_1, \lambda_2)
    \pprob(\lambda_1) d\lambda_1.
\end{equation}
In general, this also works if $\lambda_1$ and $\lambda_2$
    are sets of parameters, rather than individual parameters.
One can reduce the dimensionality of a model by integrating over
    the possible choices for a subset of the parameters
    which describe their set of hypotheses.

A counting experiment can be constrained as a Poisson Point Process,
    with a likelihood representing the possibility that an observed event count
    ($N$) can occur,
    relative to the expected value of that event count ($\mu$)
    according to a model.
The Poisson likelihood is given by:
\begin{equation}
\label{eq:poisson}
\pprob(N|\mu) = \frac{\mu^{N} e^{-\mu}}{N!}.
\end{equation}
Through Stirling's approximation, one can arrive at the following expression
    for large numbers of sources:
\begin{equation}
\label{eq:sterling}
\mathrm{ln}P(N|\mu) \approx N - \mu + N \times \mathrm{ln}(\mu/N).
\end{equation}

The Inhomogeneous Poisson Point Process extends the treatment
    of a counting experiment,
    describing the likelihood for both the event rate,
    and the properties of a set of observations,
    compared to a given model
    \citep{moller2003statistical}.
This is especially useful when our observations 
    ($\gwdets$) have parameter uncertainties characterized by
    a source likelihood:
\correction{
\begin{equation}\label{eq:ell}
\gwL_j(\BinaryParameters) = \pprob(\gwdetj|\BinaryParameters, \FormationParameters) \; .
\end{equation}
Let the indices $j$ and $k$ refer to a sample binary from an
    observed (or injected) and modeled population respectively.
}
The Inhomogeneous Poisson Point Process likelihood
    is the probability, $\pprob(\gwdets|\FormationParameters)$,
    that the set of individual detections $\gwdetj$ could be observed
    under an assumed population model characterized by some assumptions $\FormationParameters$.
\begin{equation}
\label{eq:ihpp}
    \pprob(\gwdets|\FormationParameters) = \frac{\mu^{N}_{\FormationParameters}}{N!}
        e^{-\mu_{\FormationParameters}}
        \prod\limits_{j} \Big[ \int\limits_{\AllBinaryParameters}
        \gwL_j(\BinaryParameters)
        \bar{\rho}_{\FormationParameters}(\BinaryParameters)
        \mathrm{d}\BinaryParameters
        \Big],
\end{equation}
where $\mu_{\FormationParameters}$ is the expected number of observations in
    the population model characterized by $\FormationParameters$,
    $N$ is the number of experimental (or injected) observations,
    $\bar{\rho}_{\FormationParameters}(\BinaryParameters)$
    is the normalized density of the population model
    in sample parameters $\BinaryParameters$,
    and $\gwL_j(\BinaryParameters)$ is the source likelihood.

The normalized density of a particular population model
    ($\bar{\rho}_{\FormationParameters}(\BinaryParameters)$)
    is a kind of prior in $\BinaryParameters$,
    and is often represented by a sufficiently large set of samples.
We can evaluate the agreement of our observations to that model
    by integrating over that set of samples.
It is therefore useful to approximate our specific realization
    of a continuously varying density function
    using samples from our binary evolution model
    with the Dirac Delta Function:
\begin{equation}
\label{eq:dirac}
\bar{\rho}_{\FormationParameters}(\BinaryParameters) = 
    \frac{
    \sum_k^{N_{\FormationParameters}} \delta(\BinaryParameters = \BinaryParameters_k) w_k
    }{\sum_k^{N_{\FormationParameters}} w_k},
\end{equation}
where $\FormationParameters$ describe a set of model assumptions,
    $N_{\FormationParameters}$ is the number of samples
    in a population model described by those assumptions,
    $\BinaryParameters_k$ are the parameters of a particular sample,
    and $w_k$ is the weight given to a particular sample.
In this work, we use equally weighted samples ($w_k = 1.0$).
Eq. \ref{eq:dirac} therefore simplifies to 
    $(1/N_{\FormationParameters})\sum_k \delta(\BinaryParameters = \BinaryParameters_k)$.

In practice, we calculate the logarithm of the Inhomogeneous
    Poisson Point Process likelihood,
    and have found it useful to distinguish the components of this likelihood.
The component
    which measures the number of sources is the ``rate likelihood:''
\begin{equation}\label{eq:rate-likelihood}
\mathrm{ln}\pprob(N | \FormationParameters)
    = 
    \mathrm{ln}(N - \mu_{\FormationParameters}) + 
    N \times \mathrm{ln}(\mu_{\FormationParameters}/N).
\end{equation}
The component which measures the shape of the observed
    parameter distribution is the ``shape likelihood:''
\begin{equation}\label{eq:shape-likelihood}
\sum_j \mathrm{ln} \pprob(\gwdetj | \FormationParameters) =
    \sum_j \Big[
    \mathrm{ln}(
    \sum_k^{N_{\FormationParameters}} 
    \gwL_j(\BinaryParameters_k)
    )
    - \mathrm{ln}(N_{\FormationParameters})
    \Big].
\end{equation}
The ``joint likelihood'' is then the combination of these components
    (equivalent to Eq. \ref{eq:ihpp}):
\begin{equation}\label{eq:joint-likelihood}
    \mathrm{ln} \pprob(\gwdets|\FormationParameters) =
\mathrm{ln}\pprob(N | \FormationParameters) +
    \sum_j \mathrm{ln} \pprob(\gwdetj | \FormationParameters).
\end{equation}

\end{subsection}
\begin{subsection}{Characterizing the likelihood of individual LISA-bright sources}
In general, LISA-bright sources will be described by
    many parameters including astrophysical parameters
    (such as the gravitational wave frequency: $\fgw$;
    the derivative of the gravitational wave frequency: $\dot{f}_{\mathrm{gw}}$;
    and the characteristic strain: $h$),
    as well as extrinsic parameters such as sky location
    and orientation.
In section \ref{sec:detection}, we described our process
    for constructing a likelihood estimate for each source in
    $\fgw$, chirp mass ($\mc$).
By integrating over the sample population for the \textbf{m12i} galaxy
    in these parameters,
    we effectively marginalize over the extrinsic parameters
    (including sky location and phase).

When estimating the Inhomogeneous Poisson Point Process likelihood
    in this work,
    we use a multivariate normal distribution for each source
    ($\gwL(\BinaryParameters) = \gwL(\fgw, \mc))$ of the injected 
    LISA-bright DWD population.
These multivariate normal distributions are truncated at 
    $\fgw = 0$ and $\mc = 0$.
We have obtained covariance estimates from the Fisher
    information about each source (see Section \ref{sec:detection}).
In practice, these covariances are often fine enough that if we use
    them for our source likelihood model,
    it can be unlikely that any samples from the test population
    land where the likelihood for a particular source is
    substantially greater than zero.

In order to test the agreement of these populations effectively,
    we employ two strategies, we refine the likelihood model for our population
    in a couple of ways:
First, the contribution of any one source
    to the shape likelihood is cut off below 
    $\mathrm{ln}\pprob(\gwdetj | \FormationParameters) = -10$.
This prevents a stray binary in a strange part of parameter space 
    from dominating the likelihood for the population.
Second, the variance of each multivariate normal distribution
    is inflated (when appropriate) according to Scott's rule for multivariate
    density estimation \citep{scotts-rule}:
    $\sigma_{\mathrm{Scott}} = n^{-1 / (d + 4)}$
    where $n$ is the number of samples and $d$ is the number of dimensions.

Scott's rule has been used historically to 
    construct Kernel Density Estimates (KDEs) from a set of samples,
    where a multivariate normal distribution is constructed to estimate the
    contribution of each sample to the probability density.
For now, we are not truly constructing a KDE, as
    we evaluate the likelihood on a source-by-source basis,
    our number of samples is determined by our Galactic population,
    the correlation between parameters is preserved,
    and because Scott's rule only overrides the covariance estimates
    for which a smaller covariance was predicted by the Fisher information.
Our likelihood model is a Gaussian Mixture Model (GMM),
    in the sense that it is sensitive to
    individual modes of the population (one for each source).

\end{subsection}
\begin{subsection}{A one-dimensional recovery of $\alphaCE$ from an injected population}

\begin{figure}
\includegraphics[width=3.375 in]{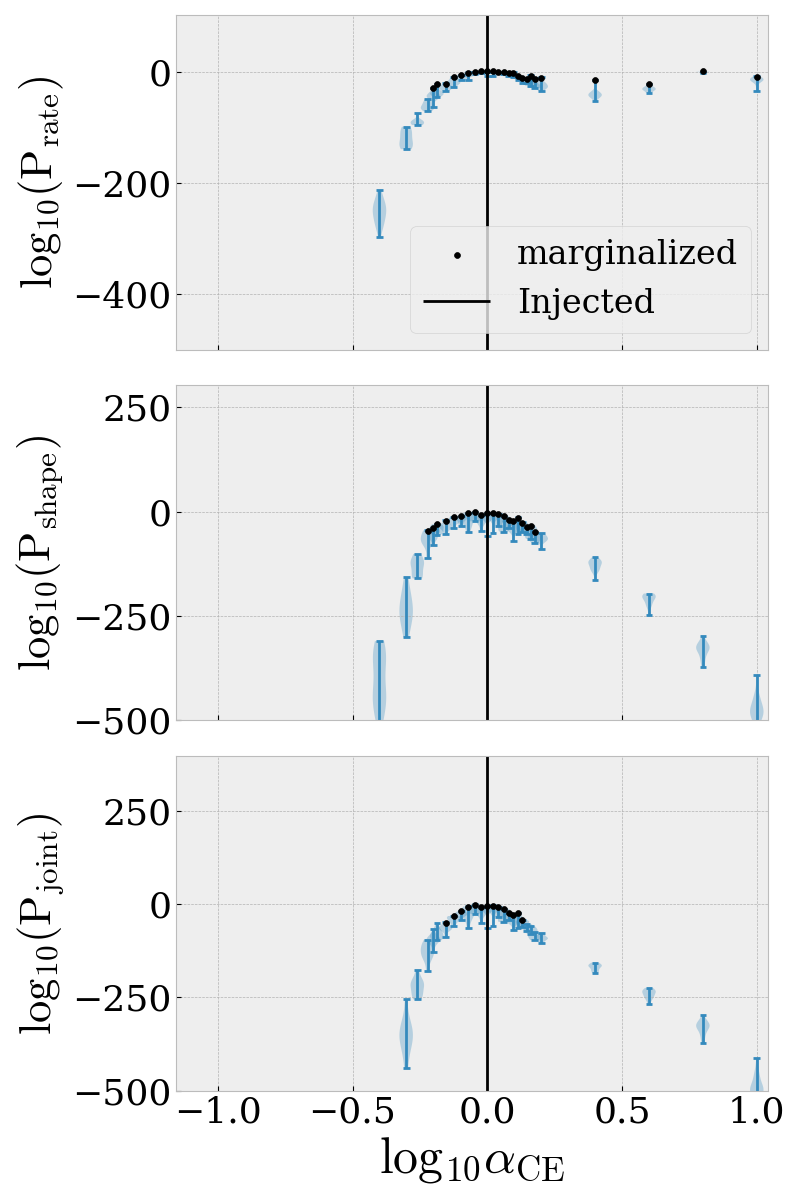}
\caption{\label{fig:injection}
    \textbf{The recovery of $\alphaCE=1.0$ in a wide view:}
    By comparing the injected LISA-bright sources with
    test populations,
    we attempt to recover the injected $\alphaCE$ parameter
    using the Inhomogeneous Poisson Point Process likelihood model.
The likelihood for several galaxy realizations for the each $\alphaCE$
    value are shown in violin plots.
Models are sampled sparsely in a logarithmic space where 
    $\alphaCE \in [0.1,10.0]$,
    and linearly in $\alphaCE \in [0.5, 1.5]$.
Both sets of models are drawn here.
}
\end{figure}

In this section,
    we explore the one-dimensional family of synthetic galaxies described in
    Section \ref{sec:1d}, in which $\alphaCE$ is the only formation parameter
    ($\FormationParameters$) we vary (see \ref{tab:CEmodels} for the
    $\alphaCE$ value assumptions).

For a given value of $\alphaCE$,
    we have up to 10 realizations of the \textbf{m12i}
    synthetic galaxy
    with different random number generator seeds.
The differences between each realization include
    both the rotation of
    star particles (about the center of \textbf{m12i})
    and the sampling of DWD binary populations.
These differences are also formation parameters,
    and can be marginalized over by taking
    the average of likelihood values for all
    realizations with a given $\alphaCE$.
In practice, this marginalization favors
    the realization with the highest log likelihood
    (this is a mean in likelihood, not log-likelihood).
Sampling from a uniform prior in the space of possible realizations
    allows us to minimize the impact of star formation sampling in
    our synthetic galaxy.
This marginalization enables us to study the impact of
    formation parameters of interest without increasing
    the dimensionality of our effective
    formation parameter space.

For the LISA-bright DWD binaries,
    we're not recording a number of instantaneous events in time,
    but we are still interested in counting the total number
    of sources identified in an injected population. 
In this work, we assume that the Inhomogeneous Poisson Point Process
    likelihood can be used to measure the agreement of an injected
    population of LISA-bright DWD binaries with test populations.
We use the Inhomogeneous Poisson Point Process likelihood
    (Eq. \ref{eq:ihpp})
    to evaluate the agreement of our injected sources
    with test populations.

By evaluating the agreement of our injected population
    at $\alphaCE = 1.0$
    with each component binary of our test populations,
    we construct the likelihood in the space of $\alphaCE$
    (see Figure \ref{fig:injection}).
In this example, we see that the rate likelihood alone fails to
    effectively constrain $\alphaCE$;
    there is a local maximum near the injected value,
    but similar rates are observed far from that peak.
However, the shape and joint likelihood of the population
    seem to be unimodal on the wide scale given here
    (ranging several orders of magnitude in $\alphaCE$).

\begin{figure}
\includegraphics[width=3.375 in]{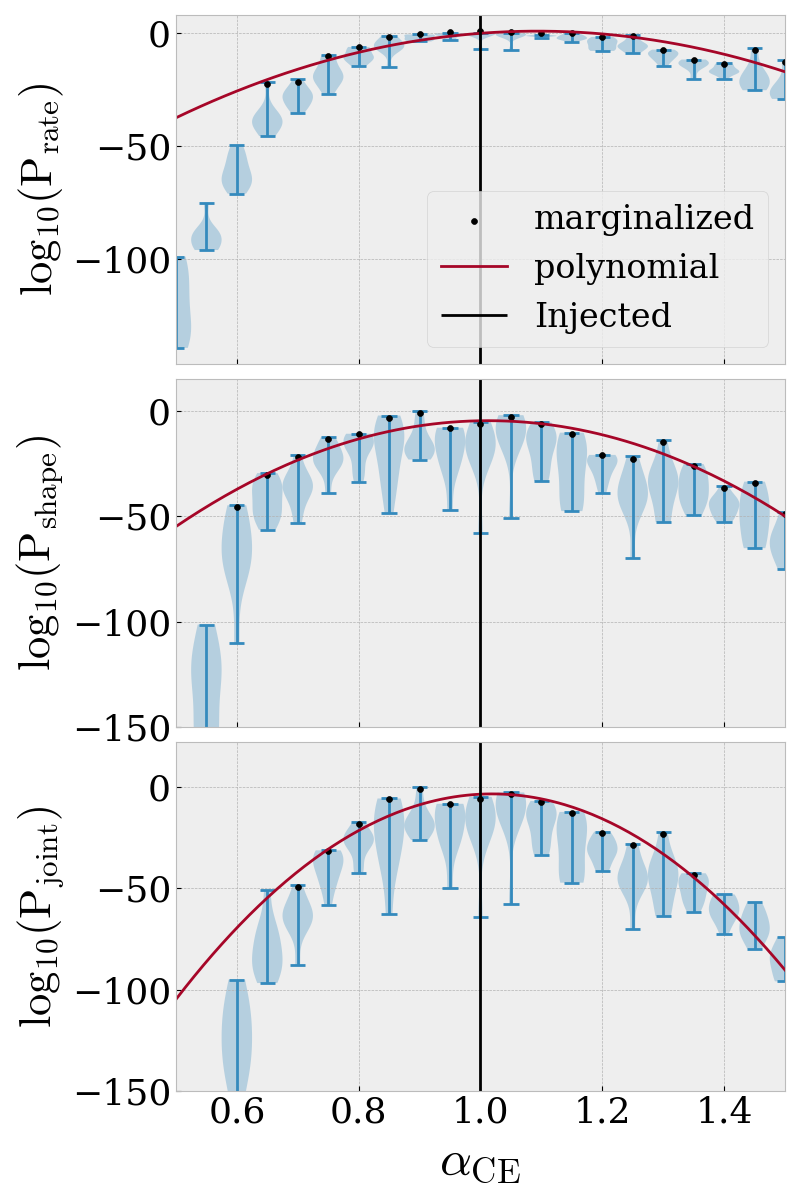}
\caption{\label{fig:true_binfrac_lnL}
    \textbf{The recovery of $\alphaCE$ in the vicinity of the peak:}
The models highlighted in this zoomed in region
    are sampled linearly (with $\alphaCE \in [0.5,1.5]$).
Each likelihood model is characterized further by a quadratic
    (in log likelihood)
    in the vicinity of the peak.
}
\end{figure}

We take a closer look at each likelihood model in the vicinity 
    of the injected value (Figure \ref{fig:true_binfrac_lnL}).
Each likelihood model has a relative maximum in this region.
Qualitatively, the joint likelihood seems to be the
    best predictor of the injected value.
This should not be surprising when varying only $\alphaCE$,
    as changes in the rate and shape of sample sources
    can both statistically be tied exclusively to common envelope physics.
We take a more quantitative approach in Section \ref{sec:fisher-alphaCE}.

\end{subsection}
 \section{Constructing a KL divergence statistic for a population}
\label{sec:kl}
\correction{
In order to guarantee that we are constructing a minimally biased model,
    it is useful to use more than one statistic
    to characterize the agreement of our injected population
    with test populations.
In addition to the Inhomogeneous Poisson Point Process likelihood,
    we also use the Kullback-Leibler (KL) divergence 
    \citep{KL-div}
    between two populations.
We use the KL divergence to measure the information lost by assuming that
    a set of observations ($p_*$) can describe a modeled population ($p$).
Similar to Section \ref{sec:likelihood},
    we describe a particular population by a number of observations ($\mu$ or $\mu_*$)
    and a discrete set of samples which represent a density $p(\BinaryParameters)$ or $p_*(\BinaryParameters)$.
For the KL divergence approach,
    we characterize the density function with the $\log_{10}(f_{\mathrm{GW}})$ and $\mathcal{M}_c$ 
    binary parameters ($\BinaryParameters$).
}

\subsection{The Kullback-Leibler divergence}
\correction{
Similar to the Inhomogeneous Poisson Point Process likelihood,
    the KL divergence between a tested population
    and reference population can be estimated with information
    about the rate of each population ($D_{\mathrm{KL}} (\mu_* | \mu)$,
    for a number of observations $\mu$),
    shape of each population ($D_{\mathrm{KL}} (p_* | p)$), or both
    ($D_{\mathrm{KL}}(p_*,\mu_* | p, \mu)$).
We test the compatibility of a set of injected populations ($p_*$)
    with several modeled populations ($p$).
    
The KL divergence between our injected observations ($p_*$) and
    a particular test model ($p$) is defined by the asymmetric expression:
\begin{equation}\label{eq:kl-shape}
D_{\mathrm{KL,shape}} = D_{\mathrm{KL}}(p_*|p)\equiv \int d\BinaryParameters  p_* \ln p_*/p  \; .
\end{equation}
The KL divergence between a number of injected or observed events ($\mu_*$) and
    an expected number of events ($\mu$) can be derived from 
    the Poisson distribution, and is simple to evaluate numerically:
\begin{equation}\label{eq:kl-rate}
D_{\mathrm{KL,rate}} = D_{\mathrm{KL}}(\mu_*|\mu) \equiv \mu_*-\mu  + \mu\ln(\mu/\mu_*)  \; .
\end{equation}
A measure of KL divergence informed by both rate and shape information
    must scale accordingly with $\mu_*$:
\begin{equation}\label{eq:kl-joint}
D_{\mathrm{KL,joint}} = D_{\mathrm{KL}}(p,\mu|p_*,\mu_*) \equiv D_{\rm KL}(\mu_*|\mu) + \mu_* D_{\rm KL}(p_*|p) .
\end{equation}
The KL divergence is 
    positive semi-definite (it will be zero if and only if $p=p_*$).

By constructing a functional form to represent the density of each population model,
    and integrating over the discrete set of samples which define observed population ($p_*$),
    we can define a discretized KL divergence:
\begin{equation}\label{eq:kl-shape-discrete}
D_{\mathrm{KL,shape}} = \frac{1}{\mu_*} \sum\limits_k \ln(p_*(\BinaryParameters_k)) - \ln(p(\BinaryParameters_k))
\end{equation}
where $p(\BinaryParameters_k)$ and $p_*(\BinaryParameters_k)$ are approximations
    to the density of sample binaries which make up $p$ and $p_*$ respectively
    (The construction of which is deferred to Section \ref{sec:kl-density}).
This is a Monte Carlo integration over samples from $p_*$ (indexed by $k$),
    and this sampling accounts for the factor of $p_*$ outside the logarithm in Eq. \ref{eq:kl-shape}.
}
\subsection{Characterizing the Density of LISA-bright sources}
\label{sec:kl-density}
\correction{
When carrying out the Monte Carlo integration in Eq. \ref{eq:kl-shape-discrete},
    we must construct an approximation to the sample densities that make up $p$ and $p_*$.
These approximations take the form of a Gaussian mixture model,
    where $p(\BinaryParameters) = \sum_j\gwL_j(\BinaryParameters) / \mu$ and
    $p_*(\BinaryParameters) = \sum_{j*}\gwL_{j*}(\BinaryParameters) / \mu_*$.
Here, $\gwL_j$ and $\gwL_{j*}$ are individual Gaussian components of the mixture model,
    and we construct them from the PE likelihood of each LISA-bright source
    (i.e. \ref{eq:ell}).
We make the same SNR and $f_{\mathrm{GW}}$ cuts as in Section \ref{sec:likelihood},
    and construct each population with the LISA-bright sample binaries.
However, we make a modification by changing coordinates from $f_{\mathrm{GW}}$ to 
    $\log(f_{\mathrm{GW}})$ in the construction of our density estimate.
We also further inflate covariance estimates
    to define the width of each Gaussian component by a minimum of twice Scott's rule
    \cite{scotts-rule}.

When sampling many galaxy realizations for each $\alphaCE$ value,
    we must choose a width for our Gaussian components which is wide enough
    so that the sampling error between realizations (sampled with different
    random number generator seeds) are not greater than the difference between 
    our assumptions for $\alphaCE$.
While it may be possible to construct a mathematical method for choosing
    the width of Gaussian components to overcome this sampling error,
    we chose this value by hand in this work.
The impact of individual samples from the test population on 
    the KL divergence is also limited by setting a maximum contribution 
    ($\ln p/p_* \leq 1$) for individual samples to the integration
    in Eq. \ref{eq:kl-shape-discrete}.
As both the Inhomogeneous Poisson Point Process likelihood
    and KL divergence models require inflated covariances
    to at least Scott's rule, a detailed parameter estimation likelihood model
    may be unnecessary. 
As the KL divergence is more susceptible to sampling error, 
    a likelihood-based approach may be preferable.

\begin{figure}
\includegraphics[width=3.375 in]{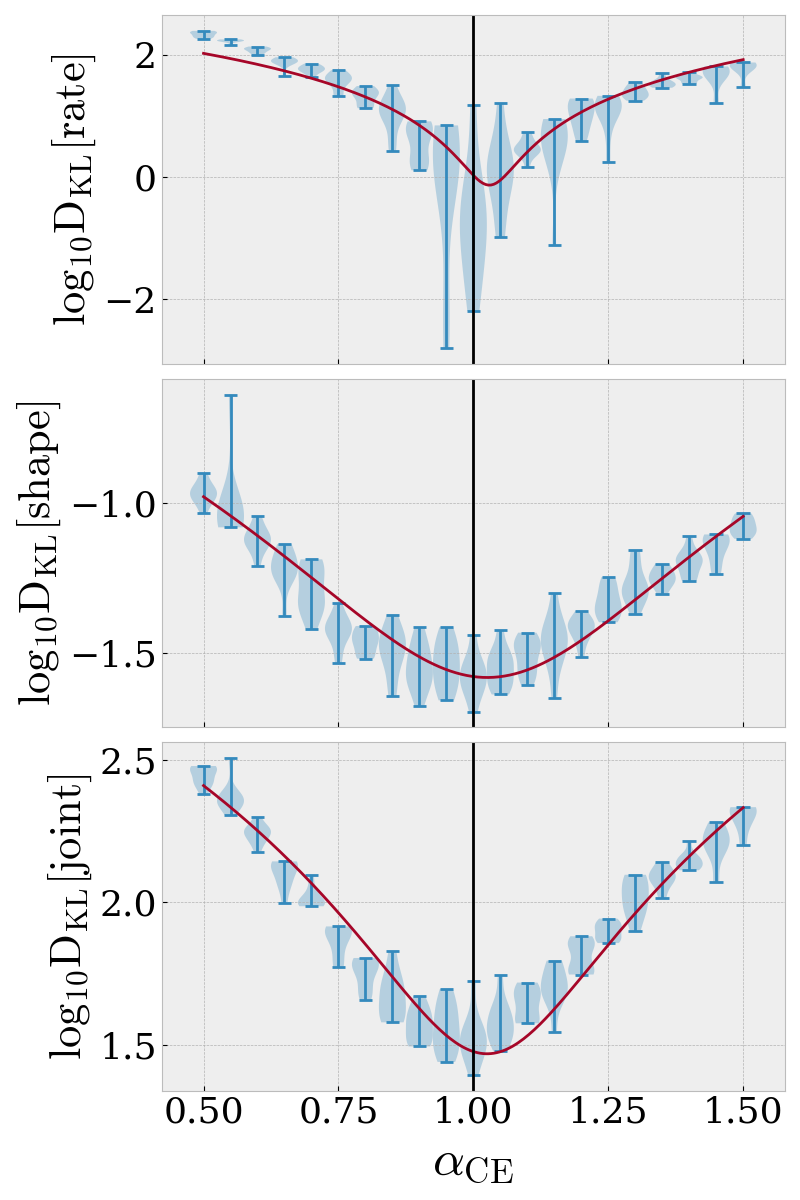}
\caption{\label{fig:true_binfrac_kl}
\textbf{The KL divergence in $\alphaCE$:}
Taking the same view as Figure \ref{fig:true_binfrac_lnL},
    we see the components of the KL divergence attributed to
    Poisson error (top), the shape of the source distribution (middle),
    and both of these together (bottom).
\correction{
We note the presence of a few realizations for which the estimated KL
    divergence appears to be numerically unstable,
    despite our choices of coordinates and inflated Gaussian
    components.
}
}
\end{figure}

As a concrete example of this KL divergence approach
    Figure \ref{fig:true_binfrac_kl} shows the KL divergence between our Gaussian mixture models
    representing the injected CEb60 population and test populations with different 
    $\alphaCE$.
In the top panel,
    we show the KL divergence calculated using only information about the expected number of events.
In the second panel, 
    we compute the KL divergence derived using only the DWD orbital parameter distributions $p(\BinaryParameters)$.
In the third panel, we show an overall combination:
In all three cases, the minimum value of the KL divergence occurs (as expected) 
    near the true parameter value, even
    allowing for different galaxy realizations. 
}

\subsection{A relationship between likelihood and KL divergence}
\label{sec:agreement}
\begin{figure}
\includegraphics[width=3.375 in]{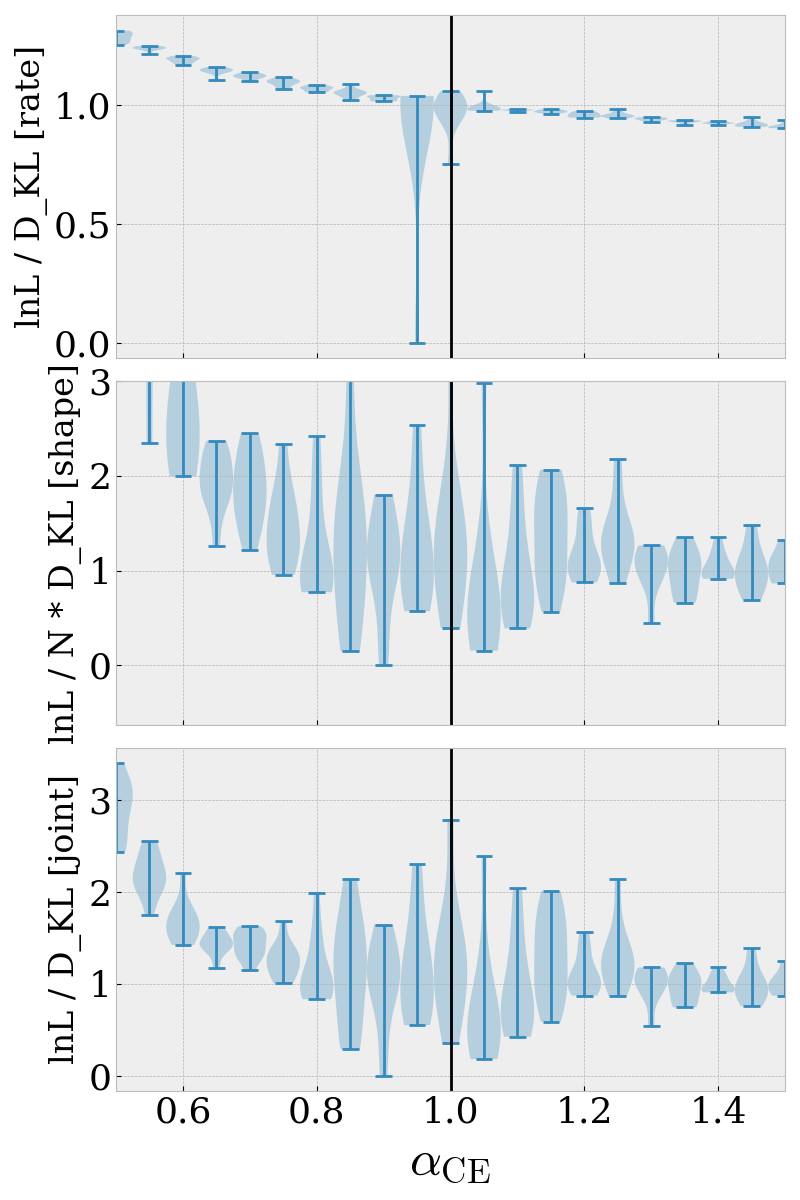}
\caption{\label{fig:true_binfrac_criteria}
    \textbf{Agreement of the KL divergence and log likelihood:}
Taking the same view again as Figure \ref{fig:true_binfrac_lnL},
    we measure the agreement of two methods of constraint:
    the log likelihood and KL divergence.
\correction{
As per Eq. \ref{eq:1204.3117.14}, these values should be close to 1.0
    if our model is well-constructed.
We note that this is essentially true in the vicinity of the
    injected model ($\alphaCE = 1.0$).
}
}
\end{figure}

\correction{
To understand why the KL divergence and likelihood provide similar information,
    we review their expected relationship.
This section is heavily inspired by \citet{1204.3117}.

Just as we are interested in finding the \emph{more likely} model
    when using a maximum likelihood approach 
    (rather than fixating on the value of a particular likelihood estimate),
    we are more interested here in the difference between KL divergence
    values for particular models
    (rather than the value of a particular KL divergence estimate).
Any proposed realization of a ``bootstrapped'' reference population
    that approximates the samples belonging to $p_*$
    will predict on average $\mu_*$ observations,
    with a distribution $p_*$, consistent with the injected population model.
Following \citet{1204.3117},
    the difference in KL divergence between the bootstrapped population and a given
    test population (characterized by $\mu$ observations and a distribution of $p$)
    are intimately related to the (expected) likelihood difference between
    the tested population model and injected population model.
Averaging over possible realizations of the
    bootstrapped source population model, we find
\begin{equation}
\E{\ln P(\{d_j\}|\Lambda)} = \E{\ln P(n|\mu) } + \E{n} \int d \BinaryParameters p_*(\BinaryParameters) \ln p(\BinaryParameters) \; .
\end{equation}
where we note $\E{n}=\mu_*$.

Reorganizing, we can express the difference in (expected) log likelihood between any two
candidate populations $(\mu_A,p_A)$ and $(\mu_B,p_B)$ characterized by $\FormationParameters_{\mathrm{A}}$
    and $\FormationParameters_{\mathrm{B}}$ in terms of differences in KL divergences:
\begin{eqnarray}\label{eq:1204.3117.14}
- \Big<\mathrm{ln} \pprob(\gwdets|\FormationParameters_{\mathrm{A}}) 
    - \mathrm{ln} \pprob(\gwdets|\FormationParameters_{\mathrm{B}})\Big> = \\ \nonumber
    [
    \mathrm{D}_{\mathrm{KL}}(\mu_*|\mu_{\mathrm{A}}) -
    \mathrm{D}_{\mathrm{KL}}(\mu_*|\mu_{\mathrm{B}})
    ] + \\ \nonumber
    \mu_* [
    \mathrm{D}_{\mathrm{KL}}(p_*|p_A) -
    \mathrm{D}_{\mathrm{KL}}(p_*|p_B)
    ]
\end{eqnarray}
In the special case that $B$ is our mixture model approximating the (injected) reference population,
    then the likelihood difference will be zero if
    and only if $A$ is also the bootstrapped reference population.
Critically, we can therefore check if an estimated KL divergence and likelihood estimate
    agree with our understanding of their construction by dividing the left half of this equation
    by the right.
If our model is well-constructed, the resulting expression should be consistent with unity.

Figure \ref{fig:true_binfrac_criteria} validates this relationship,
    showing the close agreement between the rate
    likelihood and the Poisson KL divergence $D_{\mathrm{KL,rate}}(\mu_*|\mu)$;
    the shape likelihood and the shape-only KL divergence
    $\mu_* D_{\mathrm{KL,shape}}(p_*|p)$ ;
    and the overall likelihood and the ``joint'' KL divergence introduced above.

The key differences between this approach and the Inhomogeneous Poisson Point Process likelihood
    are:
\begin{itemize}
\item The KL divergence is sensitive to a difference between two models,
    rather than the agreement of a single model to a fixed set of observations 
    (compare Eq. \ref{eq:kl-shape} and Eq. \ref{eq:ihpp}).
\item 
    In the likelihood approach, we only needed to construct a Gaussian mixture model
        for the injected population (not each test population).
    In the KL divergence approach,
        $p$ and $p_*$ are both evaluated using
        the functional form approximation (i.e. the bootstrapped model)
        from their own individual population samples.
\end{itemize}
}

 \section{How well is an injection recovered?}
\label{sec:goodness}
\begin{subsection}{Estimating the Fisher uncertainty about formation parameters}
\label{sec:fisher-alphaCE}
\begin{table}[!ht]
\centering
\begin{tabular}{|c|c|c|c|}
\hline
Method & Component & $\E{\alphaCE}$ & $\delta \alphaCE \approx \sqrt{\Gamma_{\alpha \alpha}}$ \\
\hline
$\mathrm{ln}\mathrm{P}$ & Rate & 1.093 & 0.096 \\
\hline
$\mathrm{ln}\mathrm{P}$ & Shape & 1.012 & 0.072 \\
\hline
$\mathrm{ln}\mathrm{P}$ & Joint & 1.019 & 0.052 \\
\hline
$\mathrm{D}_{\mathrm{KL}}$ & Rate & 1.030 & 0.052 \\
\hline
$\mathrm{D}_{\mathrm{KL}}$ & Shape & 1.026 & 0.053 \\
\hline
$\mathrm{D}_{\mathrm{KL}}$ & Joint & 1.025 & 0.035 \\
\hline
\end{tabular}
\caption{\label{tab:goodness}
\textbf{How well our common envelope assumption is recovered:}
For the injected parameter ($\alphaCE = 1.0$), these are our
    Fisher uncertainties in an injection recovery study though the 
    likelihood and KL divergence methods, considering
    the event rate and/or shape of the distribution.
}
\end{table}

A classic estimate for the measurement uncertainty of model hyperparameters 
    $\Lambda$ follows from a second-order
    expansion of the log-likelihood about its local maximum.  
In our context,
    a second-order expansion of the inhomogeneous Poisson log likelihood
    about the true parameters $\Lambda_*$ produces an estimate of
    $\Gamma_{ab}$,
    the inverse covariance of the posterior distribution over the model hyperparameters $\Lambda$:
\begin{equation}
\ln P(\{d_j\}|\Lambda) \simeq \ln P(\{d_j\}|\Lambda_*)  - \frac{1}{2} \Gamma_{ab}(\Lambda_a -\Lambda_{a,*})(\Lambda_b -\Lambda_{b,*}).
\end{equation}
For example, using the local polynomial approximations presented in 
    Section \ref{sec:likelihood}, 
    we can estimate $\Gamma_{\alpha\alpha}$ for the full likelihood;
    see, e.g., Figure \ref{fig:true_binfrac_lnL}.  
Conversely, exploiting the duality between $\ln P$ and $D_{KL}$ established in 
    Section \ref{sec:kl}, 
    we can also estimate $\Gamma_{ab}$ from a quadratic-order approximation to 
    the KL divergence \citep[see Eq. 15 of][]{1204.3117}.

Table \ref{tab:goodness} exhibits the agreement of quadratic fits to
    the likelihood and KL divergence for each method with
    the injected value.
Both methods provide a recovery the injected formation assumption
    for $\alphaCE$,
    and the expected $\alphaCE$ value for each method
    agrees with the true value within $10 \%$,
    and with each other method within the predicted error range
    (where $\delta \alphaCE$ is estimated by the Fisher value: 
    $\sqrt{\Gamma_{\alpha \alpha}}$).

\end{subsection}

 \section{Discussion}
\label{sec:discussion}
In this work we have assessed the impact of common envelope evolution 
    uncertainties on the population of DWDs that LISA may discover. 
We simulated 480 populations of DWDs across 15 metallicities and
    32 common envelope ejection efficiencies using COSMIC.
We then applied the metallicity-specific star formation history of The Milky-Way-like
    galaxy \textbf{m12i} from the FIRE-2 Latte Simulation Suite
    to create a synthetic Milky-Way-like population of DWDs. 
With this synthetic population, we investigated LISA's ability to distinguish between models for 
    the common envelope ejection efficiency using the measured strain and 
    gravitational wave frequency evolution. 
Finally, we explored the potential for an Inhomogeneous Poisson point process
    likelihood to make those distinctions and compared with predictions
    made using a KL divergence approach,
    through both the number and properties of predicted detections.

We summarize our results as follows:
\begin{itemize}
    \item The common envelope ejection efficiency, $\alphaCE{}$, 
        directly correlates with the \emph{number} of DWDs that 
        form per unit solar mass.
        \correction{
        As $\alphaCE$ increases, the number of 
        stellar mergers from failed common envelope ejections
        among the progenitors of DWD binaries decreases.
        }
    \item The effects of changing the common envelope ejection efficiency 
        are most prominently seen in the lowest mass stellar progenitors 
        which produce He WD binaries since these systems are the most 
        susceptible to failed common envelope ejections and subsequent 
        stellar mergers.
    \item 
        \correction{
    The effects of changing $\alphaCE$ mentioned above
        are also seen in the LISA Galactic DWD and LISA bright populations;
    however, the LISA bright population is less effected due to
        competition with the DWD confusion foreground.
        }
\item
    Both the Inhomogeneous Poisson point process likelihood and
        KL divergence statistics can be used to quantitatively
    \correction{
    distinguish between different models for DWD formation.
We find that because the KL divergence method is more susceptible
        to sampling error, we reccommend the Inhomogenous Poisson Point Process
        likelihood for future population studies.
}
    \item 
    \correction{
    Because inflated Gaussian covariances which are larger than PE uncertainties
        are required to overcome sampling error,
        we find that applying Scott's rule makes the Fisher estimates unnecessary.
}
    \item 
    \correction{
    The strongest Bayesian inference constraints are found when constraining formation
    }
        models with both the
        number and properties of LISA observable DWD systems.
        \emph{The predicted number of detections alone may not be sufficient to
        distinguish between models.}
\item 
    \correction{
    When considering only changes to common envelope ejection efficiency,
        we find that LISA is sensitive at about the 10 \% level.
        Future studies should consider how constraints change when
            incorporating other uncertain models
            (E.g. different common envelope models, mass transfer stability).
    }
\end{itemize}    

By demonstrating a recovery of the underlying astrophysical assumptions
    in a one-dimensional population model,
    we have demonstrated a method of learning about the isolated binary evolution
    formation channel for close DWD binaries.
Based on our recovery of the common envelope efficiency parameter,
    we conclude that common envelope physics are a good target for
    future studies of binary population synthesis studies
    when real LISA observations are recorded.
We also demonstrate a method of using
    Bayesian hierarchical inference methods popularized for current studies of LVK
    observations to constrain LISA DWDs.

In addition to the bright DWD population,
    LISA may observe a population of neutron star white dwarf binaries
    \citep{Nelemans2001,Moore2023,Korol2023NS},
    stellar mass black hole binaries during inspiral
    \citep{Sesana2016},
    and stochastic gravitational wave Galactic foreground
    \citep{Karnesis2021}.
It may be possible to learn about the formation
    and evolution of binary stars from these binaries as well.
Future work may explore the impact of these potential sources
    on population inference.

By studying various formation assumptions for the LISA Galactic DWD binaries,
    we not only prepare ourselves for future
    investigations of LISA observable populations,
    but we also create tools for constructing synthetic populations of LISA DWD binaries
    for a given choice of assumptions for use in other contexts.
One such example is the LISA global fit pipeline
\citep{TysonGlobalFit2023,GaoGlobalFit2023,FinchGlobalFit2023,
    LackeosGlobalFit2023,KatzGlobalFit2024,StrubGlobalFit2024}.
Many such pipelines have been using the same DWD populations for a long time.
In collaboration with these efforts our work can both provide simulation tools 
    to support DWD inference pipeline and catalog development and can begin 
    prototyping end-to-end hierarchical inference analyses needed to 
    better understand the physical processes which produce DWDs.
 
\begin{paragraph}{Data/Code availability}
A limited set of data supporting the findings of this study is openly available
    at \url{https://zenodo.org/records/13872540}\citep{DelfaveroZenodo24a}.
\correction{
The code supporting the findings of this study is available
    at \url{https://zenodo.org/records/14421368}\citep{DelfaveroZenodo24b}.
}
\end{paragraph}

\begin{acknowledgments}
The authors thank Ann E. Hornschemeier 
\correction{
and the anonymous referee for useful proofreading and feedback.
}
VD is supported by an appointment to the NASA Postdoctoral Program at the NASA Goddard Space Flight Center administered by Oak Ridge Associated Universities under contract NPP-GSFC-NOV21-0031.
ROS  gratefully acknowledges support from NSF awards NSF PHY-1912632, PHY-2012057, PHY-2309172, AST-2206321, and the Simons Foundation.
JB is supported by the NASA LISA Project Office.

We acknowledge software packages used in this publication,
    including
    NUMPY \citep{harris2020array}, SCIPY \citep{2020SciPy-NMeth}, 
    MATPLOTLIB \citep{Hunter_2007}, 
    ASTROPY \citep{astropy:2013,astropy:2018,astropy:2022}, 
    H5PY \citep{collette_python_hdf5_2014},
    LEGWORK \citep{LEGWORK},
and ldasoft \cite{ldasoft}.
This research was done using resources provided by the 
    Open Science Grid \citepOSG,
    which is supported by the National
    Science Foundation awards \#2030508 and \#1836650,
    and the U.S. Department of Energy's Office of Science. 
\end{acknowledgments}

\bibliography{Bibliography.bib}
\bibliographystyle{aasjournal}

\end{document}